\documentclass[final,authoryear,5p,times,twocolumn]{elsarticle}
\usepackage{epsfig}
\usepackage{amssymb}

\journal{Journal of Theoretical Biology}

\begin{document}

\begin{frontmatter}

\title{Selfishness, fraternity, and other-regarding preference in spatial evolutionary games}

\author[mfa]{Gy\"orgy Szab\'o}
\author[mfa]{Attila Szolnoki}

\address[mfa]{Research Institute for Technical Physics and Materials
Science, P.O. Box 49, H-1525 Budapest, Hungary}

\begin{abstract}
Spatial evolutionary games are studied with myopic players whose payoff interest, as a personal character, is tuned from selfishness to other-regarding preference via fraternity. The players are located on a square lattice and collect income from symmetric two-person two-strategy (called cooperation and defection) games with their nearest neighbors. During the elementary steps of evolution a randomly chosen player modifies her strategy in order to maximize stochastically her utility function composed from her own and the co-players' income with weight factors $1-Q$ and $Q$. These models are studied within a wide range of payoff parameters using Monte Carlo simulations for noisy strategy updates and by spatial stability analysis in the low noise limit. For fraternal players ($Q=1/2$) the system evolves into ordered arrangements of strategies in the low noise limit in a way providing optimum payoff for the whole society. Dominance of defectors,  representing the ''tragedy of the commons'', is found within the regions of prisoner's dilemma and stag hunt game for selfish players ($Q=0$). Due to the symmetry in the effective utility function the system exhibits similar behavior even for $Q=1$ that can be interpreted as the "lovers' dilemma".
\end{abstract}

\begin{keyword}
evolutionary games \sep social dilemmas \sep selfishness \sep fraternity 

\end{keyword}

\end{frontmatter}

% \linenumbers

\section{Introduction}
\label{introduction}

The evolutionary game theory has been evolving and expanding progressively in the last years due to the emergent experimental facts and the deeper understanding of models developed [for a survey see the books by \citet{nowak_06}, \citet{sigmund_10}, and reviews by \citet{szabo_pr07} and by \citet{perc_bs10}]. In the last decades huge efforts are focused on the emergence of cooperative behavior because of its importance in many human and biological systems \citep{camerer_03,gintis_10}. In the first multi-agent evolutionary systems the repeated interaction is described by the payoff matrix of traditional game theory \citep{neumann_44} and the evolution is governed by a dynamical rule resembling the Darwinian selection \citep{maynard_82}. The systematic investigations have explored the relevance of the games itself (as interactions including the set of strategies), the connectivity structure, and also the dynamical rule. Recently, the co-evolutionary games have extended the original frontiers of evolutionary games by introducing additional (personal) features and complex dynamical rules allowing the simultaneous time-dependence in each ingredient of the mathematical model. The possible personal character of players can be enhanced further in a way to postulate players who consider not just their own payoffs but also the neighbors' income. 

To elaborate this possibility we will study the social dilemmas with players located on a square lattice and collect income from $2 \times 2$ games played with all their nearest neighbors. Now, it is assumed that the myopic players wish to maximize their own utility function when they adopt another possible strategy. Throughout this utility function the players combine the self-interest with the other-regarding preference in a tunable way. Besides it, the applied strategy adoption rule involves some noise (characterizing fluctuations in payoffs, mistakes and/or personal freedom in the decision) that helps the system evolve towards the final stationary state via a spatial ordering process.  

The present work was motivated by our previous study considering the consequence of pairwise collective strategy updates in a similar model \citep{szabo_pre10b}. It turned out that the frequency of cooperators is increased significantly in the case of Prisoner's Dilemma (PD) when two randomly chosen players favors a new strategy pair if it increases the sum of their individual payoff. The latter strategy update can be interpreted as a spatial extension of cooperative games where players can form coalitions to enhance the group income. Some aspects of the other-regarding preference is modelled very recently by \citet{wang_z_pa11} who studied a spatial evolutionary PD game with synchronized stochastic imitation. On the other hand, the experimental investigations of the human and animal behavior \citep{flood_ms58,milinski_prsb97,fehr_n02,fehr_n03b} have also indicated the presence of different types of mutual helps \citep{mitteldorf_jtb00}, like charity \citep{li_yx_pa10}, inequality aversion \citep{fehr_qje99,bolton_aer00,bolton_aer06,scheuring_jtb10,scheuring_bs10,bo_x_pa10}, and juvenile-adult interactions \citep{lion_tpb09}. 

The above-mentioned relevant improvement in the level of cooperation has inspired us to quantify the effect of the group size and the number of players choosing new strategies simultaneously. From a series of numerical investigations we could draw a general picture that can be well exemplified with the present simpler model. More precisely, the most relevant improvement is achieved for those cases where each myopic player has taken into consideration all of her neighbor's payoff together with her own payoff with equal weight when selecting a new strategy. Now we will study a more general model where the utility function of each player is combined from her own payoff and her co-players' payoff with weight factors $(1-Q)$ and $Q$. In this notation $Q=0$ represents a selfish myopic player who wish to maximize her own personal income irrespective of others. The fraternal players with $Q=1/2$ favor to optimize the income (redistributed and) shared equally between each pair when choosing another strategy. The present model allows us to investigate the effect of other-regarding preference in spatial models for different levels of altruism ($Q$). For the most altruistic case ($Q=1$) the players wish to maximize the co-players' income and the system behavior also exhibits a state resembling the "tragedy of the commons" that can be interpreted as the "lovers' dilemma".

It is emphasized that the resultant formalism (payoff matrix) of the other-regarding preference was already investigated by \citet{taylor_c_e07} as a model to capture the kin- and group-selection mechanisms \citep{taylor_c_tpb06,wild_jtb07}. The origin of the basic idea goes back to the work of \citet{maynard_82} and \citet{grafen_ab79} who studied animal behaviors between relatives. The present work can be considered as a continuation of the mentioned investigations. Now our attention will be focused on the consequences of structured population for a myopic strategy update.

Using the terminology of social dilemmas \citep{dawes_arp80,santos_pnas06,roca_plr09} the above mentioned model with a stochastic myopic strategy update is defined and contrasted with other versions in the following section. The results of Monte Carlo (MC) simulations are summarized in Section \ref{mc} for a finite noise level. Section \ref{stabanal} is addressed to the spatial stability analysis in the zero noise limit and the main results are discussed in the final section.

\section{Model}
\label{model}

\subsection{Brief overview}
\label{overview}

The possible solutions of the two-person, two-strategy evolutionary games depend on the details including the spatial structure, the range of interactions, the payoffs, the dynamical rule(s), the payoffs, and the measure of noise \citep{szabo_pr07,roca_plr09,chen_xj_pre09,perc_njp09,liu_rr_epl10,dai_q_njp10,rong_pre10}. Before specifying the presently studied model accurately we briefly survey the main features of these solutions. 

We consider a simple model with players located on the sites $x$ of a square lattice (consisting of $L \times L$ sites with periodic boundary conditions). Each player can follow one of the two strategies called unconditional cooperation (${\bf s}_x=C$) or unconditional defection (${\bf s}_x=D$) within the context of social dilemmas. If these strategies are denoted by two-dimensional unit vectors, as
\begin{equation}
\label{eq:strat}
{\bf s}_x = C= \left( \matrix{1 \cr 0 \cr }\right) \;\; \mbox{and}\;\;
{\bf s}_x = D= \left( \matrix{0 \cr 1 \cr }\right) \;,
\end{equation}
then the payoff of player $x$ against her neighbor at site $x + \delta$ can be expressed by the following matrix product:
\begin{equation}
\label{eq:tpo}
P_x= {\bf s}^{+}_x {\bf A} \cdot {\bf s}_{x + \delta },
\end{equation}
where ${\bf s}^{+}_x$ denotes the transpose of the state vector ${\bf s}_x$. The payoff matrix is given as
\begin{equation}
{\bf A}=\left( \matrix{ 1 & S \cr
                       T & 0 \cr} \right)\;, 
\label{eq:pom}
\end{equation} 
where the reward of mutual cooperation is chosen to be unity ($R=1$) and the mutual defection yields zero income ($P=0$) for both players without any loss of generality. The cooperator receives $S$ (sucker's payoff) against a defector who gets $T$, the temptation to choose defection. This terminology was originally introduced for the description of Prisoner's Dilemma (PD) (where $T>R>P>S$) and later extended for weaker social dilemmas \citep{dawes_arp80,santos_pnas06}, too. 

\begin{figure}[ht!]
\centerline{\epsfig{file=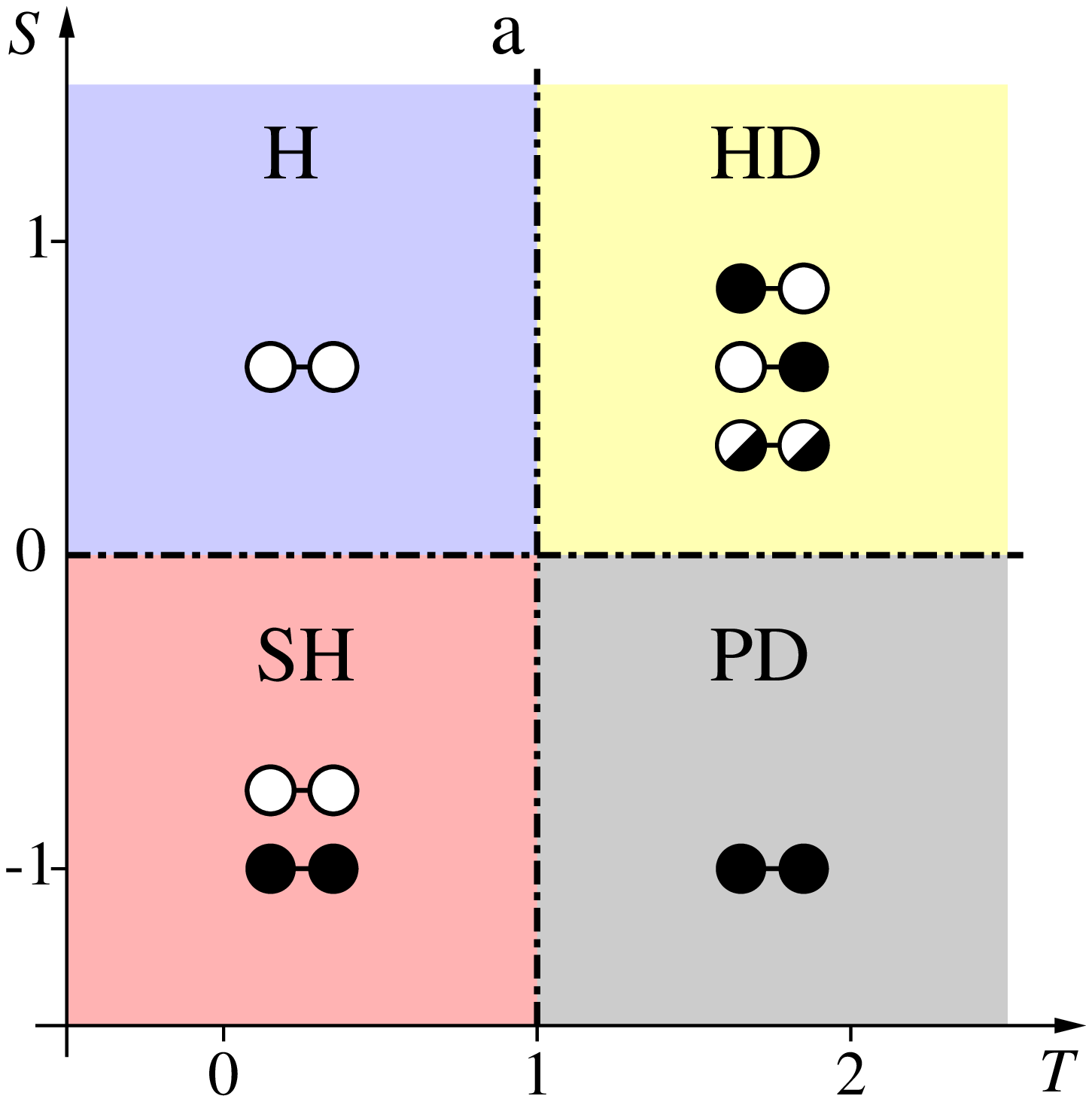,width=4.1cm}  \epsfig{file=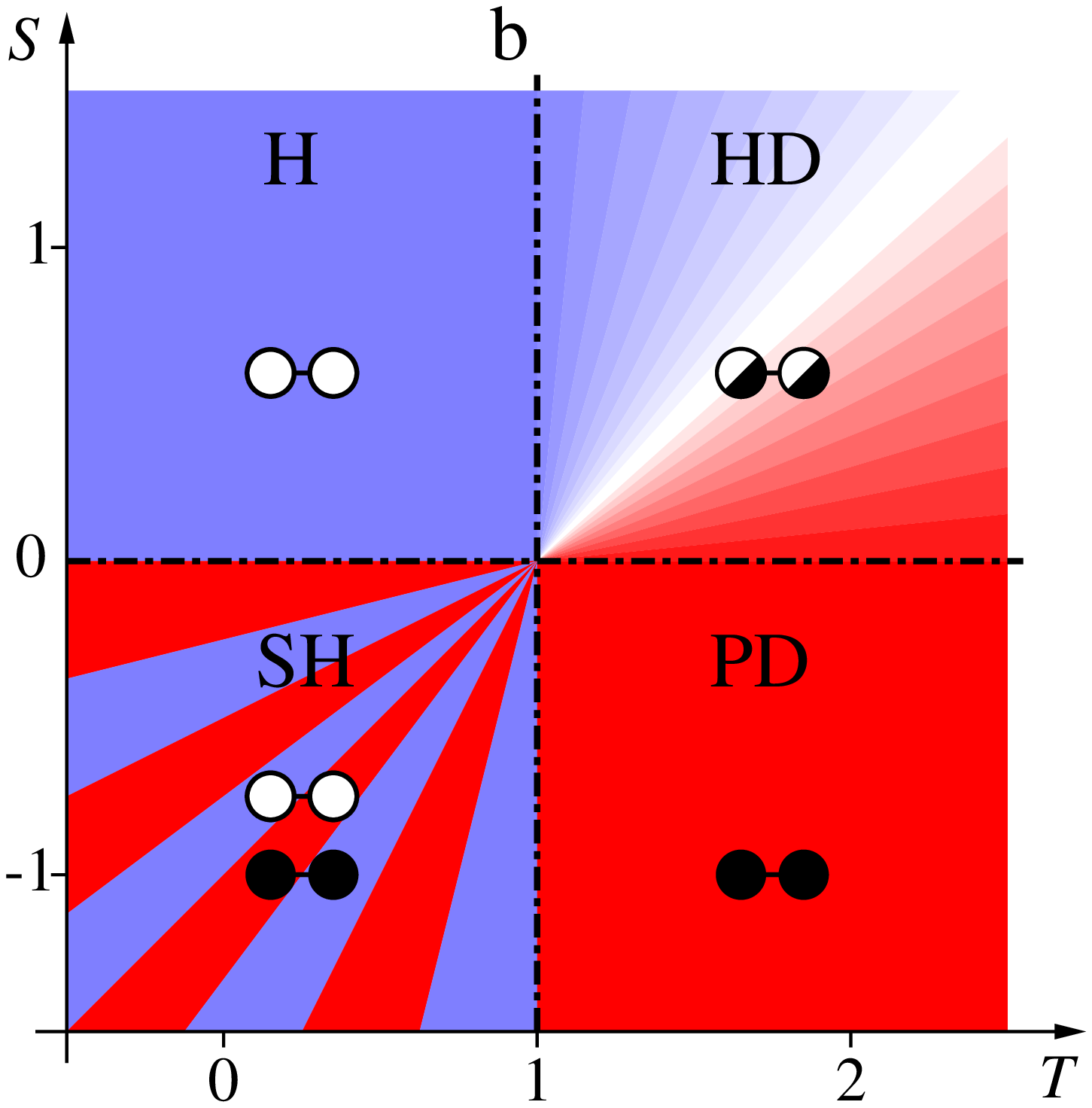,width=4.1cm}}
\vspace{0.1cm}
\centerline{\epsfig{file=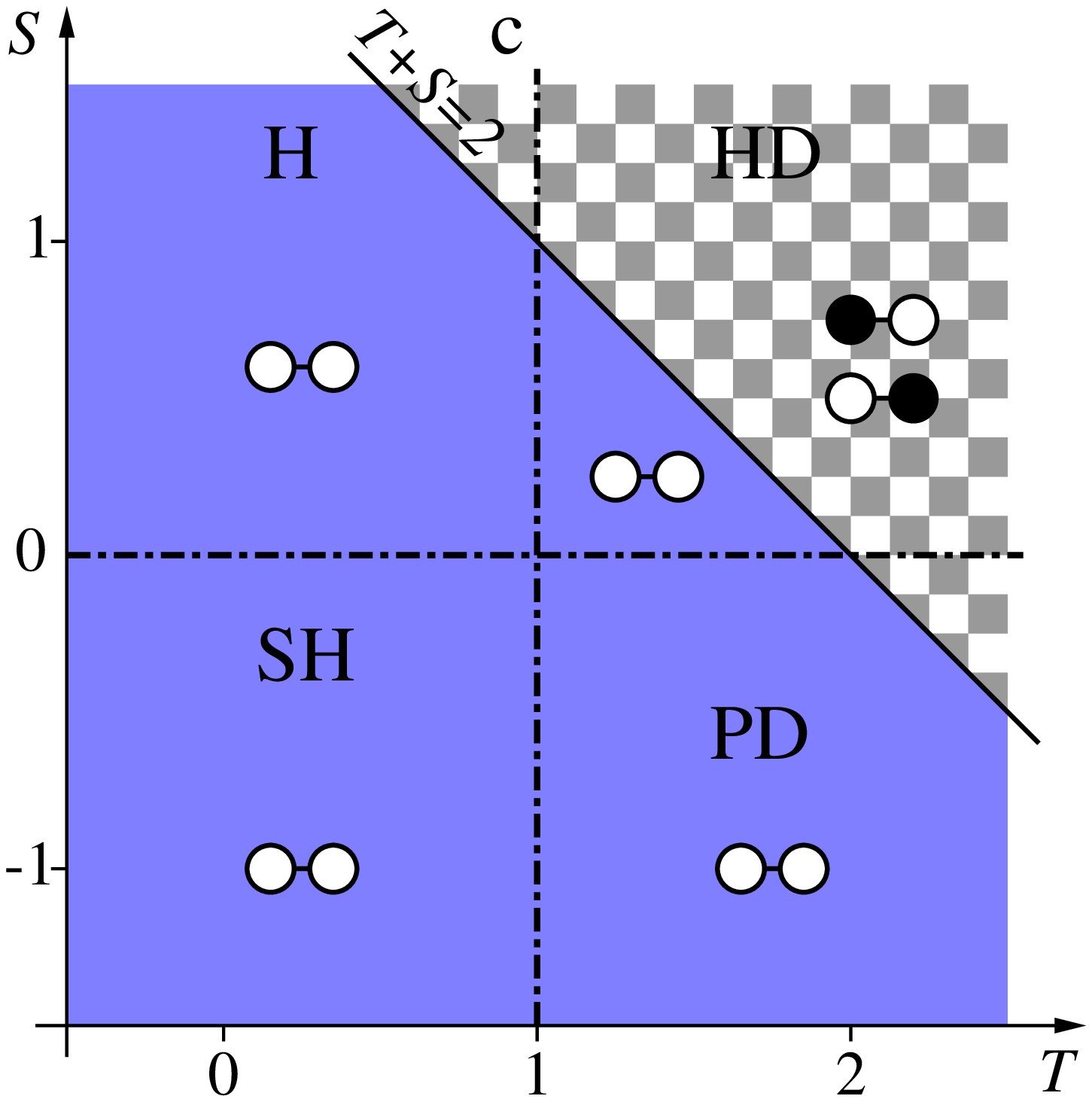,width=4.1cm}  \epsfig{file=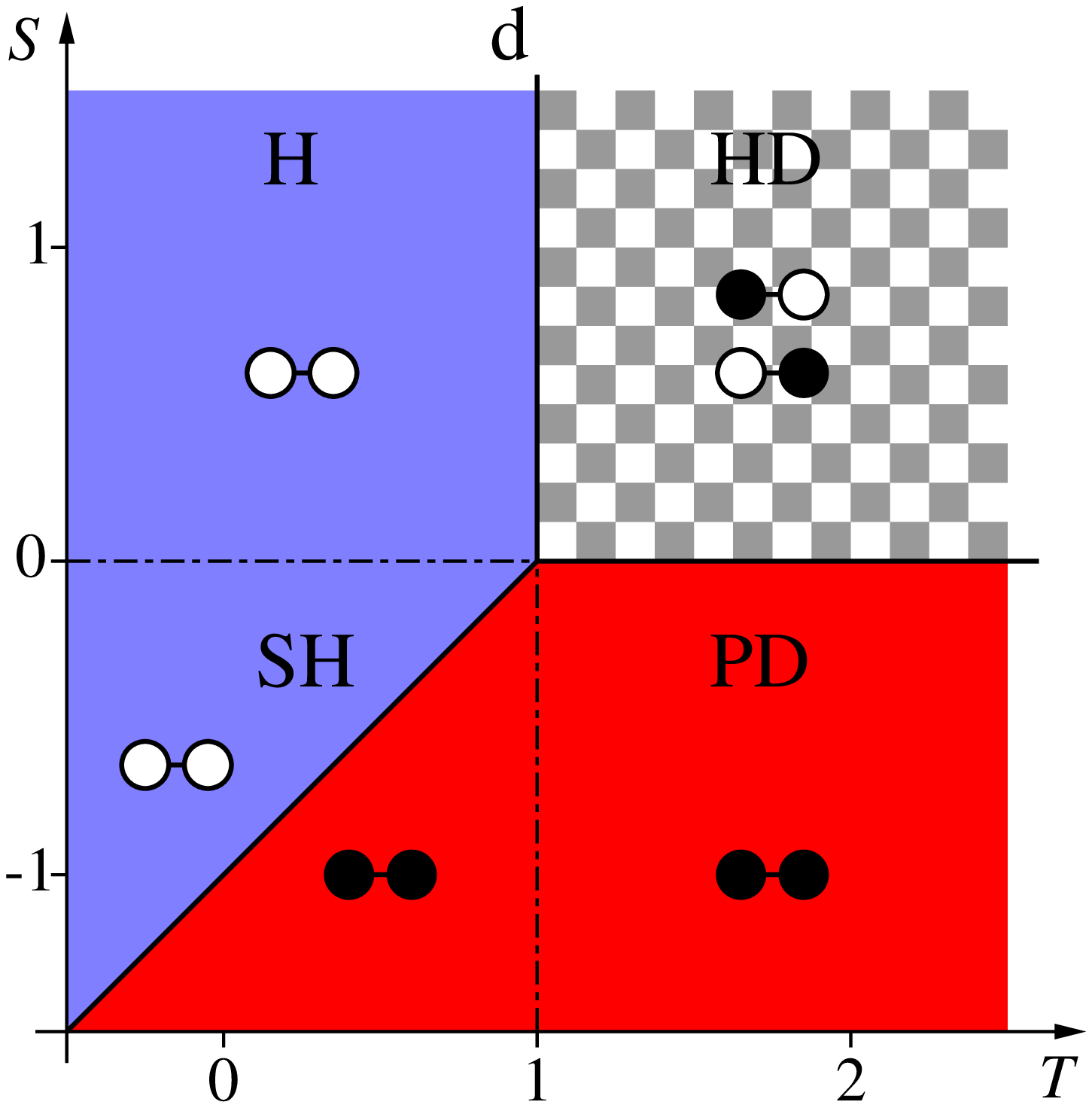,width=4.1cm}}
\caption{(Color online) Overview of possible solutions on $T-S$ plane for different conditions. (a) Nash equilibria for two-person one-shot games. (b) Stable strategy pairs for replicator dynamics in well-mixed population. (c) Optimal strategy choice on the square lattice with nearest-neighbor interactions and (d) the results of myopic strategy update in the zero noise limit for the same connectivity structure. The chessboard like patterns refer to similar arrangement of cooperators and defectors on the square lattice for nearest-neighbor interactions. The regions of Harmony (H), Hawk-Dove (HD), Stag Hunt (SH) games, and Prisoner's Dilemma (PD) are separated by dashed-dotted lines. The symbols refer to Nash equilibria (a), evolutionarily stable strategies (b), and strategy pairs on neighboring sites (c and d) where the empty, closed, and half-filled circles indicate cooperative, defective, and mixed strategies.}
\label{fig:socdil}
\end{figure}

For this notation the $S-T$ payoff plane can be divided into four segments (see Figs.~\ref{fig:socdil}) where the possible Nash equilibria are denoted by pairs of open, closed, or half-filled circles referring to pure cooperation, pure defection, and mixed strategies, respectively. For example, within the range of Harmony (H) game the Nash equilibrium (from which no player has incentives to deviate unilaterally) is the (Pareto optimal) mutual cooperation. On the contrary, for the Prisoner's Dilemma (PD) the only Nash equilibrium is the mutual defection yielding the second worst payoff. Figure~\ref{fig:socdil}a illustrates that both the $D-D$ and $C-C$ strategy pairs are Nash equilibria within the region of Stag Hunt (SH) game where $S<0$ and $T<1$. For the Hawk-Dove (HD) game besides the $D-C$ and $C-D$ strategy pairs there is an additional (so-called mixed) Nash equilibrium where the players can choose either cooperation or defection with a probability dependent on the payoff parameters. 

Many relevant features of these types of evolutionary games have already been clarified. For example, if the frequency of cooperators and defectors are controlled by a replicator dynamics (favoring the strategy of higher payoff) within a well-mixed population then the system evolves towards a final stationary state related to the possible Nash equilibria as illustrated in Fig.~\ref{fig:socdil}b \citep{nowak_06}. Namely, cooperators (defectors) die out within the region of PD (H game) while they can coexist for the case of HD game. The radial segmentation of the region of SH game in Fig. \ref{fig:socdil}b indicates that here the system develops into one of the homogeneous phase and the final result depends on the composition of the initial state.

The spatial arrangement of players (with a short range interaction), however, affects significantly the evolutionary process depending on the payoffs. For the present lattice system one can evaluate the total sum of individual payoffs for the ordered arrangement of $C$ and $D$ strategies. Figure~\ref{fig:socdil}c shows that the maximum total payoff is achieved for homogeneous cooperation ($s_x=C, \forall x)$ if $T+S<2$. In the opposite case ($T+S>2$) the total (or average) payoff is maximized if the cooperators and defectors are arranged in a chessboard-like manner as indicated by the pattern \citep{szabo_pre10b}. 

The comparison of the Figs.~\ref{fig:socdil}a and \ref{fig:socdil}c illustrates the relevant differences between the suggestions of traditional game theory when assuming two selfish players and the optimum total payoff with respect to the whole society with players located on the sites of a square lattice. Notice, that the chessboard-like arrangement of cooperators and defectors can provide optimal total payoff within a region involving a suitable part of the PD, the HD, and the H games. The curiosity of these systems is more pronounced if we compare it with the consequences of different evolutionary rules in the lattice models \citep{nowak_n92b,nowak_ijbc93,szabo_pr07,roca_plr09}. Finally, in Fig.~\ref{fig:socdil}d we summarize only the results of a spatial evolutionary game \citep{szabo_pre10b} obtained when the myopic players can choose another strategy if this change increases their own payoff assuming quenched neighborhood in the zero noise limit. 

\subsection{Model specification}

In the present work the latter dynamical rule is extended by allowing players to consider not only their own but also their neighbors' payoffs. The system is started from a random initial strategy distribution where both $C$ and $D$ strategies are present with the same frequency. The evolution of the strategy distribution is controlled by repeating the random sequential strategy updates in myopic manner. Accordingly, in each elementary step we choose a player $x$ at random who can modify her strategy from ${\bf s}_x$ to ${\bf s}_x^{\prime}$ ({\it e.g.}, $D \to C$ or {\it vice versa}) with a probability:
\begin{equation}
W({\bf s}_x \to {\bf s}_x^{\prime}) =\frac{1}{1+\exp[(U({\bf s}_x)-U({\bf s}_x^{\prime}))/K]} \;,
\label{eq:dyn}
\end{equation}
where $K$ characterizes the average amplitude of noise (that can appear for fluctuating payoffs) disturbing the players' rational decision and the utility function combines the payoffs of the player $x$ with the payoffs of her co-players within the same games. Namely,
\begin{equation}
\label{eq:util}
U({\bf s}_x)= \sum_{\delta}[(1-Q){\bf s}^{+}_x {\bf A} \cdot {\bf s}_{x + \delta }+Q{\bf s}^{+}_{x+\delta} {\bf A} \cdot {\bf s}_x]\,,
\end{equation}
where the summation runs over all nearest neighbors pairs. 
In this notation $Q$ characterizes the strength of altruism of the player. For simplicity, we now assume that all players have have the same $Q$ value, thus the whole population is described by the same attitude of selfishness. For $Q=0$ the players are selfish and the resultant behavior has already been explored in a previous work \citep{szabo_pre10b}. If $Q=1/2$ then each player tries to maximize a payoff obtained by sharing equally the common payoffs between the interacting players. In the extreme case ($Q=1$) players focus exclusively on maximizing the other's income. To give real-life example, the latter behavior mimics the attitude of lovers or the behavior of relatives in biological systems as discussed by \citep{taylor_c_e07,taylor_c_tpb06}.

Notice that the present model can be mapped into a spatial evolutionary game with selfish ($Q=0$) players for myopic strategy update if we introduce an effective payoff matrix,
\begin{equation}
{\bf A}_{eff}=\left( \matrix{1 & (1-Q)S+QT \cr
                       (1-Q)T+QS & 0 \cr} \right)\;. 
\label{eq:effpom}
\end{equation}
The effective payoff matrix becomes symmetric for $Q=1/2$ and the corresponding model is equivalent to a kinetic Ising model where the evolution of spin configuration is controlled by the Glauber dynamics in the presence of a unit external magnetic field \citep{glauber_jmp63}. In the latter case the system evolves towards a stationary state where the probability of a configuration can be described by the Boltzmann statistics and the laws of thermodynamics are valid \citep{stanley_71,blume_geb03}.

Finally we emphasize that the simultaneous exchanges $S \leftrightarrow T$ and $Q \leftrightarrow (1-Q)$ leave the system (${\bf A}_{eff}$) unchanged. This is the reason why the MC analysis can be restricted to the cases where $0 \leq Q \leq 1/2$. 

\section{Monte Carlo results}
\label{mc}

\begin{figure}[ht]
\centerline{\epsfig{file=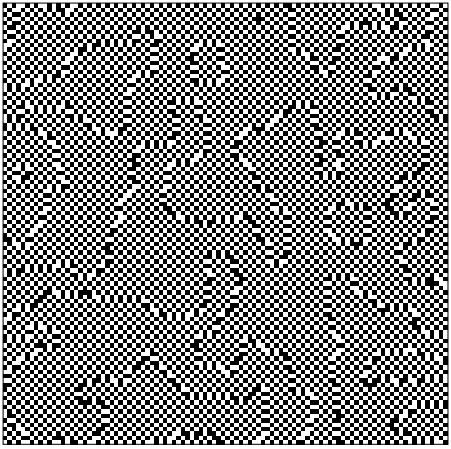,width=6cm}}
\caption{\label{fig:domgrow}The system evolves through such typical patterns into a state conquered by only one of the ordered chessboard-like arrangement. This snapshot shows cooperators (white boxes) and defectors (black boxes) on a $100 \times 100$ portion of a larger system for $T=1.5$, $S=0.5$, $Q=0$ and $K=0.25$ at time $t=30$ MCS (Monte Carlo steps) if the system started from a random initial state.}
\end{figure}

The MC simulations are performed when varying the payoffs $S$ and $T$ at a few representative values of $Q$ for $L \geq 400$ and $K=0.25$. In most of the cases the system is started from a random initial state and after a suitable thermalization time $t_{th}$ the stationary state is characterized by the fraction of cooperators ($\rho_A$ and $\rho_B$) averaged over a sampling time $t_s$ in the sublattices $A$ and $B$ corresponding to the white and black boxes on the chessboard. In fact there exist two equivalent sublattice ordered arrangements of cooperators and defectors: (1) $\rho_A \to 1$ and $\rho_B \to 0$; (2) $\rho_A \to 0$ and $\rho_B \to 1$ if $K \to 0$. During the transient time both types of ordered arrangements are present in a poly-domain structure and the typical linear size of domains growth as $l \propto \sqrt{t}$. Such a situation can occur for example within the region of HD game as demonstrated in Fig.~\ref{fig:domgrow}. 
Finally one of the ordered structure prevails the whole spatial system. Evidently, the requested thermalization time (to achieve the final mono-domain structure) increases with system size as $t_{th} \propto L^2$. 

It is emphasized that the domain growing process is blocked for those dynamical rules forbidding the adoption of irrational choices (when $U({\bf s}_x) > U({\bf s}_x)^{\prime}$). The latter cases yield frozen poly-domain patterns where $\rho_A=\rho_B$ as described by \citet{sysiaho_epjb05,wang_wx_pre06}. The present evolutionary rule allows the irrational choices with a probability decreasing very fast when $K \to 0$. Consequently, the expected value of $t_{th}$ is also increased drastically if $K \to 0$ \citep{bray_ap94}. The resultant technical difficulties can be avoided if the system is started from a biased initial state where one of the sublattice is occupied by only cooperators (or defectors) while other sites are filled randomly with $C$ or $D$ players. 

\begin{figure}[ht]
\centerline{\epsfig{file=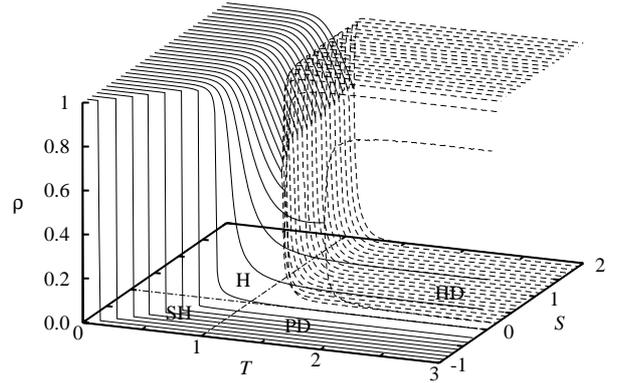,width=8cm}}
\caption{\label{fig:q00}MC results for the frequencies of cooperators in the two sublattices of the square lattice as a function of $T$ and $S$ for $K=0.25$ and $Q=0$. Dashed lines denote distinguishable $\rho$s in the sublattices $A$ and $B$ while the solid lines indicate $\rho$s when varying $T$ for a fixed $S$ in the absence of this symmetry breaking.}
\end{figure}

For selfish players ($Q=0$) the results of MC simulations are summarized in Fig.~\ref{fig:q00} in the low noise limit where the readers can distinguish three types of ordered structure. Namely, the above mentioned sublattice ordering occurs within the territory of HD game. The "homogeneous" $D$ and $C$ phases are separated by a first order transition located along the line $S=T-1$ when varying the payoff parameters. More precisely, $\rho_A=\rho_B \to 1$ in the limit $K \to 0$ if $S>T-1$ and $T < 1$. 
On the contrary, the system evolves into a state called "tragedy of the commons" ($\rho_A=\rho_B \to 0$ if $K \to 0$) when $S<T-1$ and $S < 0$. In the case of finite noise, the sharp ordered phases disappear and point defect can emerge resulting in intermediate values for cooperator frequency. Notice that the transitions from the homogeneous phases to the symmetry breaking sublattice ordered structure follow similar scenario. Namely, the stationary frequency of cooperators evolves toward a state (where $\rho_A=\rho_B \simeq 1/2$) when approaching the critical point from the homogeneous regions while $|\rho_A -\rho_b|$ vanishing algebraically from the opposite direction. The width of the transition regime in both side of the critical point is proportional to $K$. 

The sublattice ordering as a continuous transition belongs to the Ising universality class \citep{szabo_pre10b,stanley_71}. This means that the vanishing order parameter ($|\rho_A - \rho_B|$) follows a power law behavior when approaching the critical point and simultaneously the correlation length, the relaxation time, and the magnitude of fluctuations diverge algebraically. The latter effects imply an increase in the uncertainty of the numerical data. To avoid this problem we used a significantly larger size (typically $L > 4000$) and longer thermalization and sampling time ($t_{th}\simeq t_{s} > 10^6$ MCS) in the close vicinity of the transition point. Using this method we could reduce the statistical error comparable to the line thickness.  

\begin{figure}[ht]
\centerline{\epsfig{file=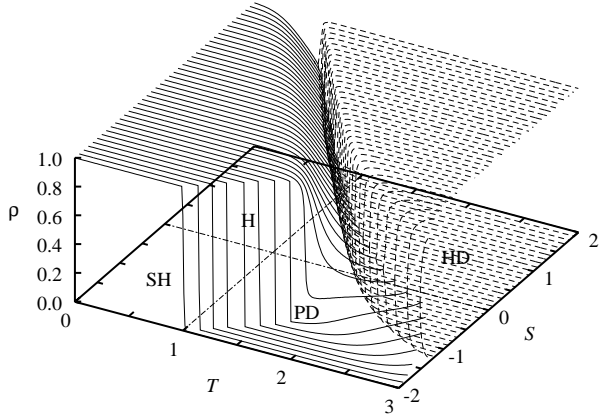,width=8cm}}
\caption{\label{fig:q033}Frequencies of cooperators in the two sublattices $A$ and $B$ as a function of $T$ and $S$ for $K=0.25$ and $Q=1/3$.}
\end{figure}

All the above mentioned three phases and also the main characteristics of the phase transitions are present for $Q=1/3$ as demonstrated in Fig.~\ref{fig:q033}. The most striking difference between Figs. \ref{fig:q00} and \ref{fig:q033} is the shift of the phase boundaries. As a consequence for $Q=1/3$ the territory of the "tragedy of the commons" (in the $T-S$ plane) is reduced. 
Similar behavior was observed in a model where the evolution is controlled by pairwise collective strategy update \citep{szabo_pre10b}.

\begin{figure}[ht]
\centerline{\epsfig{file=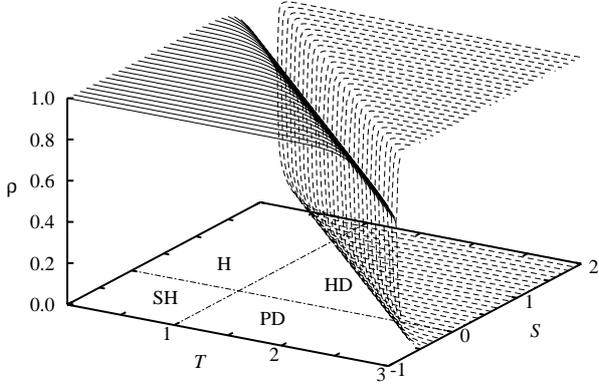,width=8cm}}
\caption{\label{fig:q05}MC results for the frequencies of cooperators in the sublattices $A$ and $B$ as a function of $T$ and $S$ if $Q=1/2$  and $K=0.25$.}
\end{figure}

Fundamentally different behavior is found for the fraternal players ($Q=1/2$) as illustrated in Fig.~\ref{fig:q05}.
In this case the results depend only on $T+S$ and the system does not fall into the state of the "tragedy of the commons". It is emphasized that in the zero noise limit the system evolves into the state providing the maximum total payoff (compare Figs.~\ref{fig:socdil}c and \ref{fig:q05}). 

The above numerical investigations were repeated for several noise levels, too. These numerical data have justified that in the zero noise limit the $T-S$ phase diagrams coincide with those we derived from stability analysis of the spatial patterns.

\section{Stability analysis}
\label{stabanal}

\begin{figure}[ht!]
\centerline{\epsfig{file=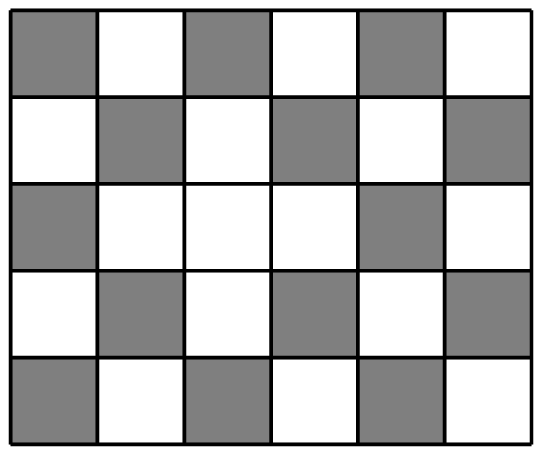,width=2.24cm}  \epsfig{file=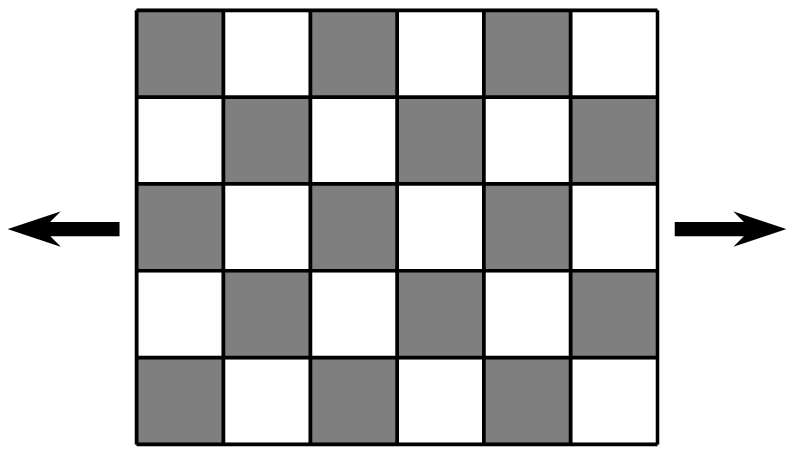,width=3.22cm}
\epsfig{file=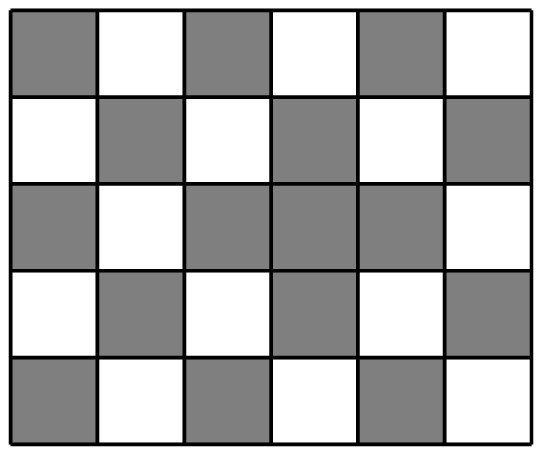,width=2.24cm}}
\caption{Possible elementary processes destabilizing the chessboard like arrangement of cooperators (white) and defectors (gray).}
\label{fig:bulkinst}
\end{figure}

To have a deeper understanding about the ordering process, we first study the stability of the sublattice ordered arrangement of strategies against a single strategy flip in the zero noise limit. The two possibilities are demonstrated in Fig.~\ref{fig:bulkinst}. 
In the first case a defector reverses its strategy if the cooperation yields higher utility, that is, if 
\begin{equation}
4[(1-Q)T+QS] < 4 
\label{eq:afdtoc1}
\end{equation}
or after some algebraic simplification
\begin{equation}
QS < 1-(1-Q)T \,.
\label{eq:afdtoc}
\end{equation}
Due to the absence of the second neighbor interactions all the defectors (within the chessboard like structure) are enforced to cooperate in the low noise limit if the condition (\ref{eq:afdtoc}) is satisfied. Consequently, in this case the sublattice ordered arrangement of cooperation and defection transforms into homogeneous cooperation. 

One can easily check that the appearance of a single defector in the state of homogeneous cooperation is favored if $QS > 1-(1-Q)T$ which is the opposite of the condition (\ref{eq:afdtoc}). The random sequential repetition of this type of point defect yields a poly-domain structure resembling to those plotted in Fig.~\ref{fig:domgrow}. As mentioned before, this poly-domain structure is not stable for $K>0$ because the fluctuations changes the sizes of ordered domains and their vanishing is not balanced by the appearance of new domains. Finally the system evolves into one of the sublattice ordered structure.     

From the above analysis one can conclude that the equation
\begin{equation}
Q(S-1) = -(1-Q)(T-1) \,
\label{eq:afcpb}
\end{equation}
determines the position of the phase boundary separating the sublattice ordered phase from the homogeneous cooperation in the $T-S$ plane. This mathematical expression reflects clearly that the straight line phase boundary rotates anti-clockwise around the point ($T=S=1$) from the vertical direction to the horizontal one when increasing the value of $Q$ from 0 to 1. 

The above described analysis can be repeated to study the stability of the sublattice ordered structure against a single strategy change from $C$ to $D$ as plotted on the right-hand side of Fig.~\ref{fig:bulkinst}. It is found that the sublattice ordered structure evolves into the homogeneous $D$ phase if $(1-Q)S < -QT$ and the resulting equation
\begin{equation}
(1-Q)S = -QT 
\label{eq:afdpb}
\end{equation}
gives a boundary line separating the sublattice ordered phase from the homogeneous defection state in the payoff plane. This phase boundary is also a straight line rotating clockwise around the point $T=S=0$ from the horizontal ($Q=0$) to the vertical ($Q=1$) direction. 

\begin{figure}[ht!]
\centerline{\epsfig{file=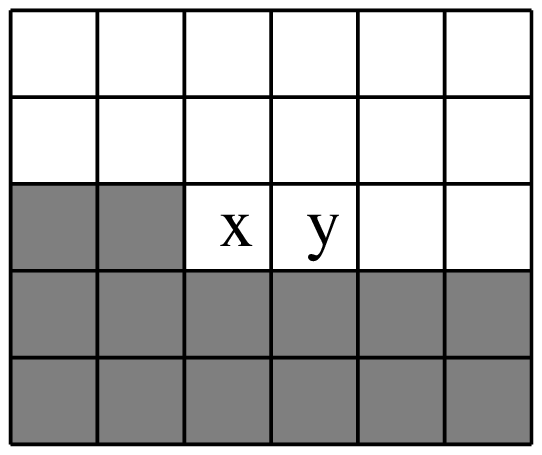,width=2.24cm}  \epsfig{file=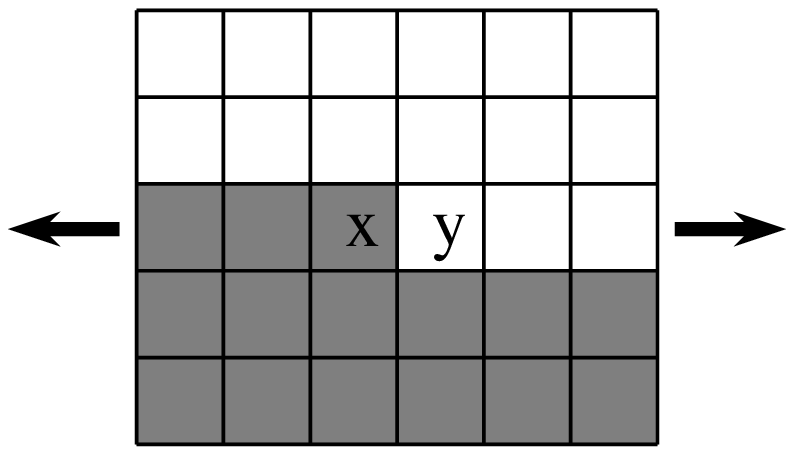,width=3.22cm}
\epsfig{file=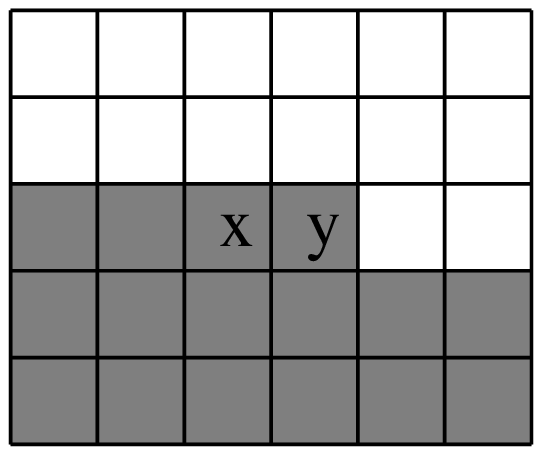,width=2.24cm}}
\caption{Elementary processes along the interface separating the cooperative (white) and defective (gray) domains of players.}
\label{fig:invasion}
\end{figure}

The phase boundaries (\ref{eq:afcpb}) and (\ref{eq:afdpb}) divide the $T-S$ plane into four segments excepting the case $Q=1/2$ when the given straight lines are parallel. For $Q=0$ these segments are equivalent to those types of games indicated in Figs. \ref{fig:socdil}. The above stability analysis allows the existence of both the homogeneous $C$ and $D$ phases within the region of SH game. The simulations indicate the presence of both phases in a poly-domain structure during a transient period. The final result of the competition between these two ordered structure can be deduced by determining the average velocity of the boundary separating the regions of homogeneous cooperation and defection. For the present dynamical rule the most frequent elementary changes are the shifts of a step-like interface as illustrated in Fig.~\ref{fig:invasion}. 
Evidently other elementary processes are also observable with a vanishing probability when $K \to 0$. The latter elementary processes (e.g. the creation of a new step) can affect only the average velocity of the interface. Using the above observation as a working hypotheses we can derive a condition for the direction of invasion that is justified by the MC simulations.

Namely, along the horizontal interface the step moves right in the zero noise limit if $U({\bf s}_x=D) > U({\bf s}_y=C)$. This yields $D$ invasion if 
\begin{equation}
2[(1-Q)T+QS] > 2[1+(1-Q)S+QT] \,.
\label{eq:dinv}
\end{equation}
In the opposite case $C$ invasion is preferred and the system evolves into the state where all the players cooperate. Thus, the position of the phase boundary separating the homogeneous $C$ and $D$ phases can be given by a straight line 
\begin{equation}
S = T - {1 \over 1-2Q} \,
\label{eq:cdpb}
\end{equation}
in the $T-S$ plane. 

Here it is worth mentioning that the phase boundary determined by (\ref{eq:cdpb}) coincides with those suggestion derived from the criterion of risk dominance \citep{harsanyi_88} favoring the selection of those strategy that provides higher utility if the co-player chooses her strategy at random. In other words, the present myopic strategy update (on the square lattice) can be considered as a realization of the criterion of risk dominance in the crucial local constellations because the players $x$ and $y$ are surrounded by two cooperators and two defectors. 

\begin{figure}[ht!]
\centerline{\epsfig{file=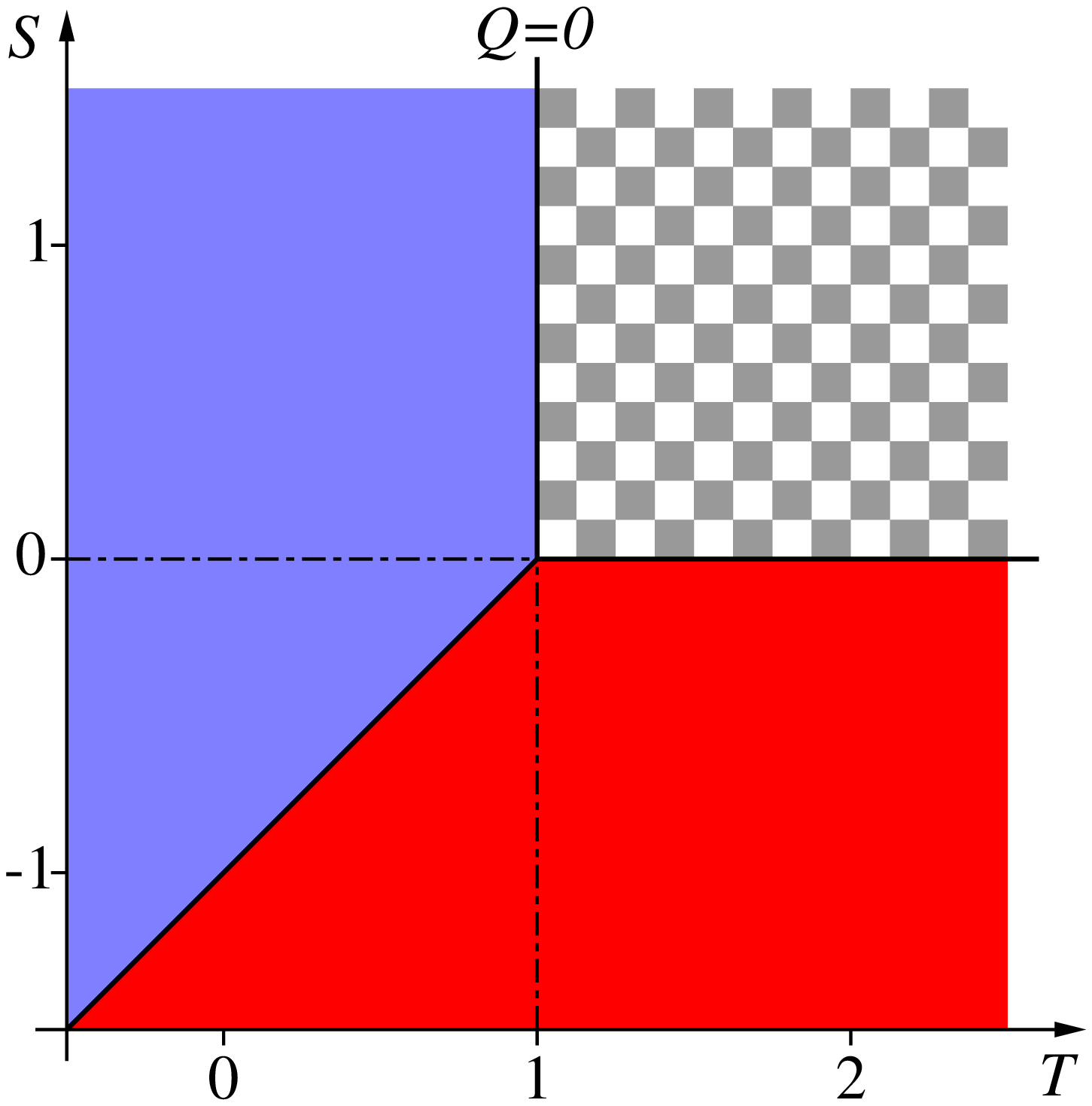,width=4.1cm}  \epsfig{file=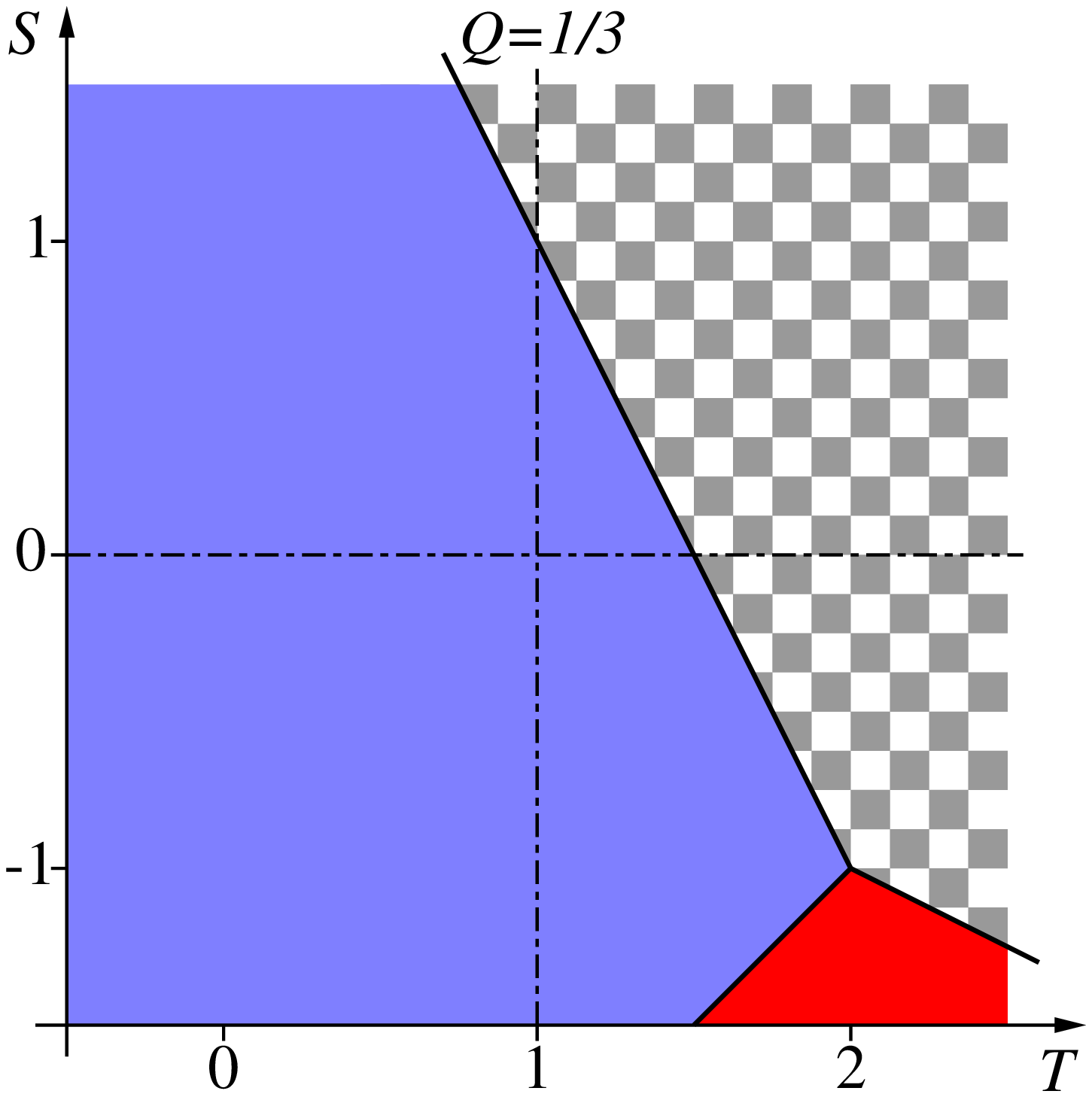,width=4.1cm}}
\vspace{0.1cm}
\centerline{\epsfig{file=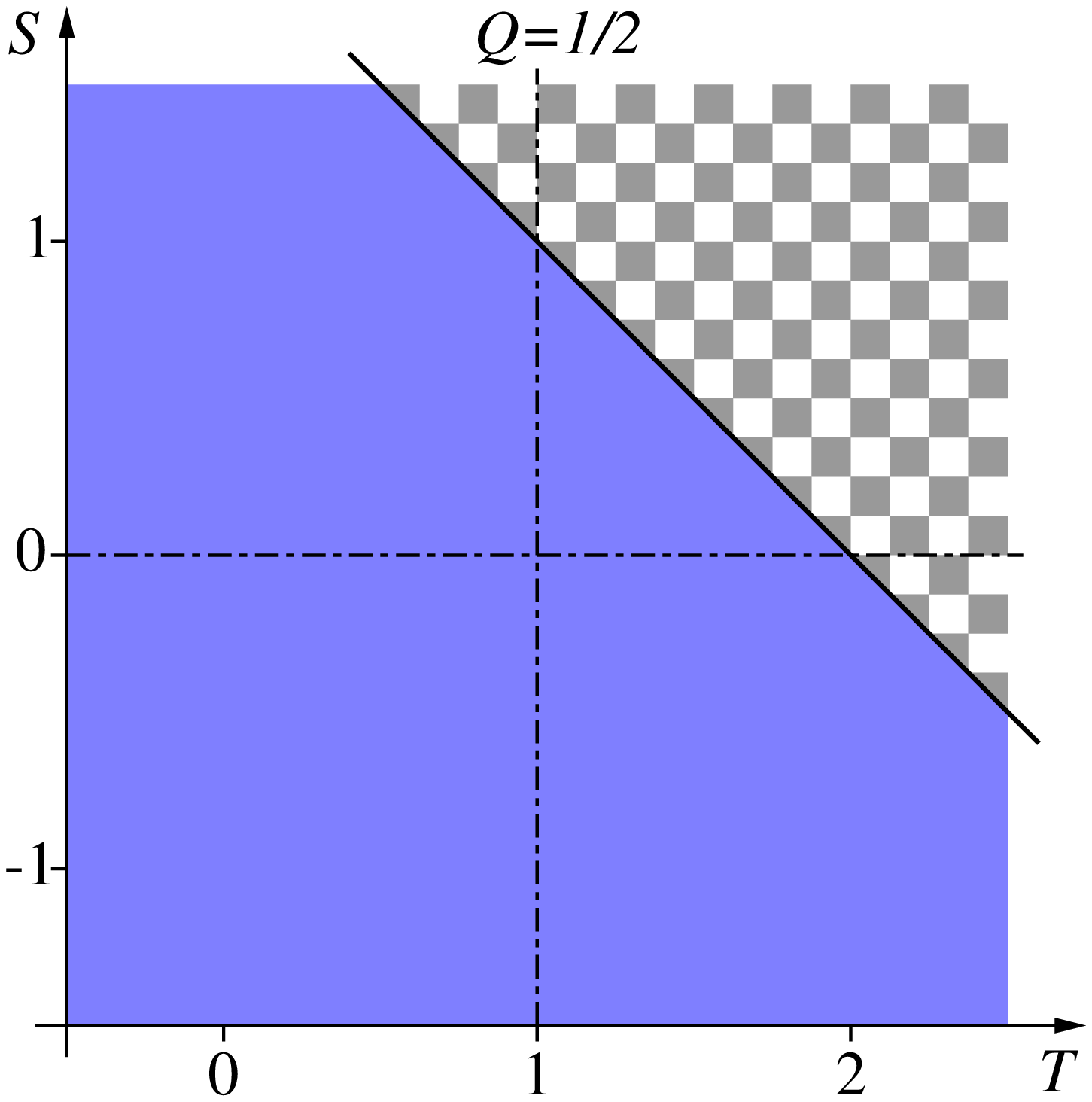,width=4.1cm}  \epsfig{file=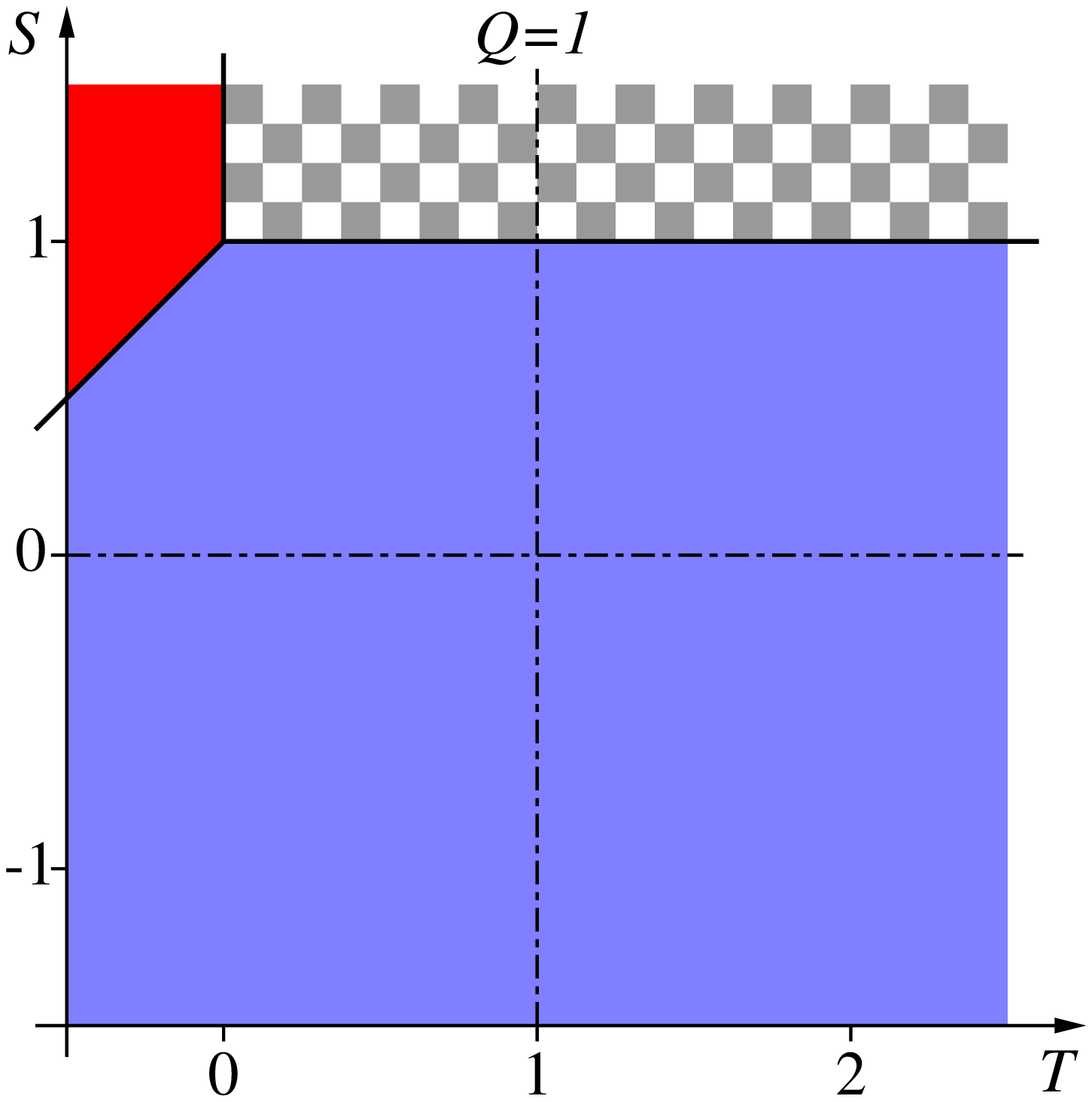,width=4.1cm}}
\caption{(Color online) $T-S$ phase diagram in the zero noise limit for four different values of $Q$. The ordered phases are denoted by the same colors as in Fig.~\ref{fig:socdil}d.}
\label{fig:multi}
\end{figure}

The results of the above stability analysis are summarized in Figs.~\ref{fig:multi}. It is emphasized that the three phase boundaries [given by eqs. (\ref{eq:afcpb}), (\ref{eq:afdpb}), and (\ref{eq:cdpb})] meet at a so-called tricritical point (if $Q \ne 1/2$) and divide the $T-S$ plane into three parts in agreement with the expectations deduced from the MC results in the low noise limit.   
Fig.~\ref{fig:multi} illustrates graphically what happens if the value of $Q$ is increased gradually. The territory of the homogeneous cooperation and the sublattice ordered structures expand in the $T-S$ plane at the expense of the homogeneous defection if $Q \to 1/2$. Evidently this process is accompanied with a relevant increase in the total payoff for the payoff parameters involved. This process is saturated for $Q=1/2$ when the system achieves the optimum total payoff for any payoff matrix. For further increase of $Q$ the process is reversed within some territory of H and SH games where the system can evolve into the "tragedy of the commons" as demonstrated in Fig.~\ref{fig:multi} for $Q=1$ when the system (effective payoff) can be mapped to the case of selfish player by exchanging the payoffs $S \leftrightarrow T$. Consequently, the overstatement of the other regarding preference may also results in a social dilemma.

\section{Summary}
\label{summary}

We have introduced a spatial evolutionary game with a myopic strategy update rule to study what happens if the players' characters, regarding the target utility, are tuned continuously. The two extreme characters are the egoist players (maximizing their own income irrespective of others) and the completely altruistic players or lovers who try to maximize the others' income irrespective of their own payoff. This feature is quantified by introducing a utility function composed from the player's and the co-player's incomes with suitable weight factors. It turned out that all the relevant results can be explained by considering an effective payoff matrix in agreement with previous investigations \citep{grafen_ab79,maynard_82,taylor_c_e07}. Despite its simplicity this model indicated clearly the importance of the fraternal behavior. Namely, the highest total income is achieved by the society whose members share their income fraternally. Any deviation from the fraternal behavior can result in the emergence of the "tragedy of the commons" within a suitable region of payoff parameters even for altruistic players characterized by the other-regarding preference. 

Finally, to exemplify the lover's dilemma we end this work by citing the opening sentences of the paper by \citet{frohlich_jcr74}: "There is a famous story written by O'Henry about a poor young couple in love at Christmas time ("The Gift of the Magi"). Neither has any money with which to buy the other a present, although each knows what the other wants. Each has only one prized possession: he, his father's gold pocket watch; she, her beautiful long hair. He has long been coveting a gold watch fob, while she has long admired a pair of tortoise shell hair combs in a nearby shop. The conclusion of the story is the exchange of gifts, along with a description of their means of getting the money for their purchases. He gives her the combs (having pawned his watch to raise the money). She gives him the watch fob (having cut and sold her hair)." The mentioned example is discussed exhaustively in the text book by \citet{barash_03}.

This work was supported by the Hungarian National Research Fund (Grant No. K-73449) and Bolyai Research Grant.

%\bibliographystyle{elsarticle-harv}
%\bibliography{egt}

\end{document}